\documentclass[pra,onecolumn]{revtex4}
\usepackage{graphicx,amsmath,amssymb,xspace}
\usepackage{color}

\newtheorem{rem}{Remark}

\begin{document}

\newcommand{\inner}[2]{\langle #1, #2 \rangle}

\newcommand{\ketbra}[2]{| #1\rangle \langle #2|}
\newcommand{\ket}[1]{| #1 \rangle}
\newcommand{\bra}[1]{\langle #1 |}
\newcommand{\Tr}{\mathrm{Tr}}
\newcommand\F{\mbox{\bf F}}
\newcommand{\h}{\mathcal{H}}

\newcommand{\PSD}{\textup{PSD}}

\newcommand{\C}{\mathbb{C}}
\newcommand{\X}{\mathcal{X}}
\newcommand{\Y}{\mathcal{Y}}
\newcommand{\Z}{\mathcal{Z}}
\newcommand{\sspan}{\mathrm{span}}
\newcommand{\kb}[1]{\ket{#1} \bra{#1}}
\newcommand{\pos}{D}

\newcommand{\thmref}[1]{\hyperref[#1]{{Theorem~\ref*{#1}}}}
\newcommand{\lemref}[1]{\hyperref[#1]{{Lemma~\ref*{#1}}}}
\newcommand{\corref}[1]{\hyperref[#1]{{Corollary~\ref*{#1}}}}
\newcommand{\eqnref}[1]{\hyperref[#1]{{Equation~(\ref*{#1})}}}
\newcommand{\claimref}[1]{\hyperref[#1]{{Claim~\ref*{#1}}}}
\newcommand{\remarkref}[1]{\hyperref[#1]{{Remark~\ref*{#1}}}}
\newcommand{\propref}[1]{\hyperref[#1]{{Proposition~\ref*{#1}}}}
\newcommand{\factref}[1]{\hyperref[#1]{{Fact~\ref*{#1}}}}
\newcommand{\defref}[1]{\hyperref[#1]{{Definition~\ref*{#1}}}}
\newcommand{\exampleref}[1]{\hyperref[#1]{{Example~\ref*{#1}}}}
\newcommand{\hypref}[1]{\hyperref[#1]{{Hypothesis~\ref*{#1}}}}
\newcommand{\secref}[1]{\hyperref[#1]{{Section~\ref*{#1}}}}
\newcommand{\chapref}[1]{\hyperref[#1]{{Chapter~\ref*{#1}}}}
\newcommand{\apref}[1]{\hyperref[#1]{{Appendix~\ref*{#1}}}}
\newcommand\rank{\mbox{\tt {rank}}\xspace}
\newcommand\prank{\mbox{\tt {rank}$_{\tt psd}$}\xspace}
\newcommand\alice{\mbox{\sf Alice}\xspace}
\newcommand\bob{\mbox{\sf Bob}\xspace}
\newcommand\pr{\mbox{\bf Pr}}
\newcommand\av{\mbox{\bf{\bf E}}}
\newcommand{\pabxy}{(p(ab|xy))}
\newcommand{\calR}{\mathcal{R}}
\newcommand{\calQ}{\mathcal{Q}}
\newcommand{\calB}{\mathcal{B}}
\newcommand{\calA}{\mathcal{A}}

\newcommand{\AMS}{\mathrm{AMS}}
\newcommand{\GHZ}{\mathrm{GHZ}}

\def\be{\begin{equation}}
\def\ee{\end{equation}}

\newcommand{\snote}[1]{\textcolor{red}{\textbf{(Jamie: #1)}}}
\newcommand{\wnote}[1]{\textcolor{blue}{\textbf{(Zhaohui: #1)}}}
\newcommand{\comment}[1]{{}}
\newcommand{\NEW}[1]{\textcolor{blue}{#1}}
\newcommand{\red}[1]{\textcolor{red}{#1}}
\newcommand{\blue}[1]{\textcolor{blue}{#1}}

\title{\vspace{-1cm} Device-independent dimension test in a multiparty Bell experiment}
\author{{Zhaohui Wei}$^{1,2,4,5,}$}\email{weizhaohui@gmail.com}
\author{Jamie Sikora$^{2,3,5,}$}\email{jamiesikora@gmail.com}
\affiliation{$^{1}$Center for Quantum
Information, Institute for Interdisciplinary Information Sciences,
Tsinghua University, Beijing 100084, P. R. China\\$^{2}$Centre for Quantum Technologies, National
University of Singapore, Singapore\\$^{3}$Perimeter
Institute for Theoretical Physics, Waterloo, Ontario,
Canada\\$^{4}$School of Physical and Mathematical Sciences, Nanyang
Technological University, Singapore\\$^{5}$MajuLab, CNRS-UNS-NUS-NTU
International Joint Research Unit, UMI 3654, Singapore}

\begin{abstract}
A device-independent dimension test for a Bell experiment aims to
estimate the underlying {Hilbert space} dimension that is required
to produce given measurement statistical data without any other
assumptions concerning the quantum apparatus. Previous work mostly
deals with the two-party version of this problem. In this paper, we
propose a very general and robust approach to test the dimension of
any subsystem in a multiparty Bell experiment. Our dimension test
stems from the study of a new multiparty scenario which we call
\emph{prepare-and-distribute}. This is like the prepare-and-measure
scenario, but the quantum state is sent to multiple,
non-communicating parties. Through specific examples, we show that
our test results can be tight. Furthermore, we compare the
performance of our test to results based on known bipartite tests,
and witness remarkable advantage, which indicates that our test is
of a true multiparty nature. We conclude by pointing out that with
some partial information about the quantum states involved in the
experiment, it is possible to learn other interesting properties
beyond dimension.
\end{abstract}

\maketitle

\section{Introduction}

Suppose we have an unknown quantum system and we want to assess its
quantum properties. One way to tackle this problem is by using only
classical information obtained by interacting with the target system
classically and thus no (possibly unrealistic) assumptions need to
be made concerning the quantum states and/or measurements involved.
For this purpose, often what people do is choose different
means/settings to measure the system, then collect the corresponding
statistical data, which is of course classical. It is well-known, on
the other hand, that if one wants to describe a quantum system
completely using only classical information, the amount of
information needed will increase exponentially with the size of the
quantum system, which is usually much more than what is collected
through measurements \cite{NC00}. Therefore, it would seem that we cannot infer
any useful information about the quantum state using a limited
amount of statistical data alone.

Interestingly, these tasks are indeed possible in some cases, and
the information inferred is said to be \emph{device-independent} \cite{Ekert91,ABM+07}.
Clearly, they are attractive not only mathematically, but also from
an application standpoint. For example, when a businessman wants to
sell a quantum product, it would help if he can convince potential
clients that the product is behaving as advertised. Instead of
taking the machine apart piece by piece and trying to convince the
buyer that there is nothing funny going on, e.g., something
maliciously entangled with his company laboratory, he can choose to
interact with it via measurements to obtain a small number of
outcome statistics, and invoke device-independent results from the
literature.

Bell experiments are typical settings to demonstrate phenomena of
device-independence \cite{BPA+08}. In such a setting, a number of spatially
separated parties share a quantum state and each party chooses one
local measurement from a finite selection to measure his/her
subsystem. The statistical data for all possible choices of
measurements is recorded as a \emph{correlation}. For bipartite
cases of Bell experiments, it has been shown that the dimension of
each party can be estimated in a device-independent manner
\cite{BPA+08,SVV15} (see also \cite{WCD08}). This problem is
motivated as follows. It is well-known that Hilbert space dimension
is a valuable resource in quantum processing tasks. Therefore, for
any quantum correlation that is generated in a quantum setting, we
often prefer the dimension required to produce this correlation to
be as small as possible. Thus, being able to estimate the underlying
Hilbert space dimension device-independently is very useful.

In particular, to solve this fundamental problem, using the fact
that some entangled quantum states can produce correlations
violating certain Bell inequalities \cite{Bell64}, the concept of \emph{dimension
witness} was proposed to estimate the underlying dimension, where
the key idea is to build a relation between dimension and the extent
that Bell inequalities are violated \cite{BPA+08}. The approach of
dimension witness requires sets of quantum correlations to be convex, thus
shared classical randomness is assumed. This approach is powerful, but it relies heavily on the availability of a Bell inequality for the statistics being tested. Assuming
that shared classical randomness is not a free resource, i.e., it is absorbed into the entangled quantum state, a new easy-to-compute dimension bound for this problem
has also been provided \cite{SVV15}. This bound is independent of any Bell
inequalities, and thus it is very convenient to use as it can be readily applied to any correlation data. Recently, the approach
of \cite{SVV15} was used to certify system dimensionality in a newly proposed
experimental platform for multidimensional quantum systems \cite{WPD+18}.
Other examples of device-independence on Bell experiments include assessing the amount of entanglement in some
bipartite cases \cite{MBL+13}, and even pinning down the underlying
quantum states completely, a task known as
self-testing~\cite{PR92,BMR92,MY98,MY04,CGS17}.

Though more than one approach has been discovered to deal with
device-independent dimension estimation of bipartite Bell
experiments, multipartite versions have not been found to the best
of our knowledge. This problem is not only important and realistic,
but also interesting in its own right as the generalization from
bipartite to multipartite cases enriches the mathematics needed
considerably as it is much more complicated. However, using the
standard approach of finding dimension witnesses based on Bell
inequalities to address this problem is a very difficult task
as this requires much knowledge of the complicated structures of
multipartite quantum correlations. Indeed, Bell inequalities in the
multiparty setting are very hard to find and are not that well
understood \cite{Svetlichny87,CGP+02}, especially compared to the two-party case. To get around
these difficulties, in this paper we develop a general technique for
this problem which results in an easy-to-compute lower bound for the
underlying dimension of any subsystem in a general multiparty Bell
experiment. To this end, we define a multiparty quantum scenario
called \emph{prepare-and-distribute}, and then propose an efficient
way to estimate the distances between quantum states in this
scenario based on measurement statistical data only. This allows us
to identify device-independently a desired lower bound for the
target dimension in the multipartite Bell setting. Through specific
examples, we show that our result can be tight. At the same time,
since we are interested in the dimensions of individual parties, in
principle we can also use methods for bipartite cases (e.g. in
Ref.~\cite{SVV15}) to tackle our problem. By a concrete example, we
illustrate that our new result in this paper is much better than
generalizations from known bipartite results. This demonstrates that
it is of a true multiparty nature. We also point out that with more
information on the target quantum state, it is possible to learn
other quantum properties beyond dimension in some circumstances.

\medskip

\section{Preliminaries}

\subsection{Multiparty Bell Scenario}

In a multiparty Bell scenario, we have $k+1$ physically separated
parties, sharing a quantum state $\rho$ acting on a $(k+1)$-partite
Hilbert space $\bigotimes_{i=1}^{k+1} \C^{d_i}$, where $d_i$ is the
dimension of the $i$-th subsystem. Each party has a local
measurement apparatus, which allows for various
measurement settings which can be applied to their subsystems. 

As not to be bound to $26$ parties, we shall call one of them Alice,
and the rest of the parties Bob-$1$, Bob-$2$, up to Bob-$k$. Alice
will have measurement settings given by a finite set $X$ and Bob-$j$
will have measurement settings from a finite set $Y_j$. Thus, when
they measure the shared quantum state $\rho$ with their chosen
settings, the probability that Alice gets outcome $a$ (from a finite
set $A$) and Bob-$j$ gets outcome $b_j$ (from a finite set $B_j$) is
given by
\begin{equation} \label{eq:multi}
p(a b_1 \cdots b_k| x y_1 \cdots y_k) = \Tr \left(\left(M^x_a
\otimes \bigotimes_{j=1}^{k} (N^j)^{y_j}_{b_j} \right) \rho \right),
\end{equation}
where $\{ M^x_a : a \in A \}$ is Alice's local positive-operator
valued measure (POVM) and $\{ (N^j)^{y_j}_{b_j} : b_j \in B_j \}$ is
Bob-$j$'s local POVM. A three-party Bell experiment is illustrated
in FIG.~1. The set of all joint conditional probabilities $p(a b_1
\cdots b_{k}| x y_1 \cdots y_{k})$ is called a
\emph{$(k+1)$-correlation} (or just \emph{correlation} when $k$ is
clear from context).

\begin{figure}[htbp]
   \label{fig:example}
   \centering
   \includegraphics[width=3.25in]{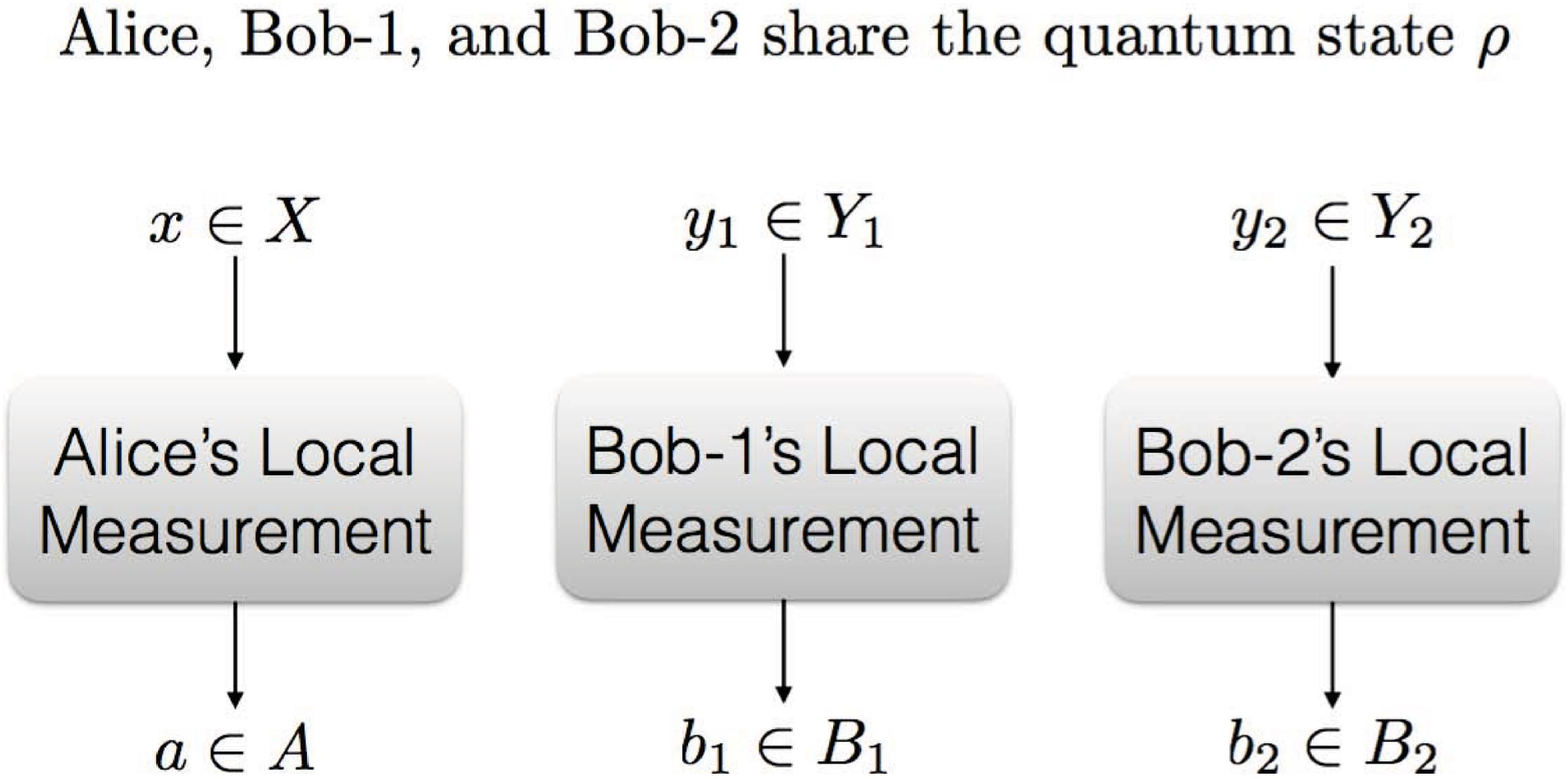}
   \caption{{Alice, Bob-$1$, and Bob-$2$ in a three-party Bell experiment.}
   }
\end{figure}

\subsection{The prepare-and-distribute scenario}

We now define a new $(k+1)$-party quantum scenario that is useful
for the purposes of this paper. Suppose a single party, say Paula,
prepares a $k$-partite quantum state $\rho_x$, for some $x \in X$,
and distributes it to $k$ different, physically separated parties,
which we call Roger-$1$, \ldots, Roger-$k$. Then Roger-$j$ measures
his corresponding subsystem with available local POVM indexed by
$y_j$ and gets the outcome $b_j$. The measurement settings and
outcomes share the same notation as in the previous discussion about
multiparty Bell experiments for reasons that will be clear shortly.
Like a $(k+1)$-party Bell correlation, a
\emph{prepare-and-distribute correlation} can be defined as below
with similar notations,
\begin{equation} \label{eq:distribute}
p(b_1 \cdots b_k| x y_1 \cdots y_k) = \Tr \left(\bigotimes_{j=1}^{k}
(N^j)^{y_j}_{b_j} \ \rho_x \right).
\end{equation}
A prepare-and-distribute experiment involving three parties can be
seen in FIG. 2. Later we will discuss the close relationship between
multiparty Bell scenarios and prepare-and-distribute scenarios.

\begin{figure}[htbp]
   \centering
   \includegraphics[width=2.25in]{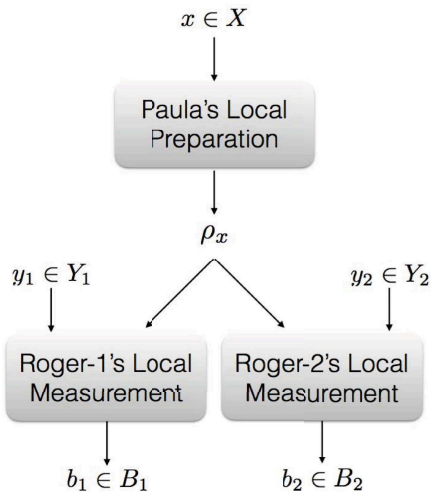}
   \caption{Paula, Roger-$1$, and Roger-$2$ in a three-party prepare-and-distribute experiment.
   }
   \label{fig:example}
\end{figure}

\section{Main results}

\subsection{Bounding distances between quantum states in a prepare-and-distribute scenario}

In this subsection, we consider the following problem: Suppose we
are given a prepare-and-distribute setting and the corresponding
correlation data $p(b_1 \cdots b_k| x y_1 \cdots y_k)$, can we give
a nontrivial estimation for the distance between two arbitrary
preparations $\rho_x$ and $\rho_{x'}$? The answer is affirmative.

In this paper, we choose the concept of \emph{fidelity} to measure
the distances between quantum states \cite{NC00}. For two quantum
states $\sigma_1$ and $\sigma_2$ acting on the same Hilbert space, their
fidelity
is defined as $\F(\sigma_1, \sigma_2) = \| \sqrt{\sigma_1}
\sqrt{\sigma_2} \|_1$. A useful property of fidelity is that for any
two quantum states, if one measures them using the same measurement,
then the fidelity between the two outcome distributions (as
classical-quantum states) is no less than that between the two
original quantum states \cite{NC00}. Therefore, by compositing all
the local POVMs on Rogers as a whole, we can immediately get an upper
bound for $\F(\rho_x,\rho_{x'})$ as below:
\begin{eqnarray} \label{eq:trivial}
\!\!\!\!\!\!\!\! \sum_{b_1 \cdots
b_k}\sqrt{p(b_1 \cdots b_k| x y_1 \cdots y_k)}
\sqrt{p(b_1 \cdots b_k| x' y_1 \cdots y_k)}.
\end{eqnarray}
If we want to optimize, we can indeed take the minimum over all
measurement settings $y_1, \ldots, y_k$ and the bound still holds.

As a crucial part of our discussion later, we  introduce a new
method to estimate $\F(\rho_x,\rho_{x'})$, which has a much better
performance than the simple bound above. To this end, we need the
\emph{expansive} property of fidelity~\cite{NC00}, which means that
\begin{equation}
\F(\Phi(\rho), \Phi(\sigma)) \geq \F(\rho, \sigma)
\end{equation}
for any quantum states $\rho, \sigma$ and any completely positive
and trace-preserving (CPTP) map $\Phi$. The CPTP map of
relevance here is the map
\begin{equation}
\rho_x \to \sum_{b_1} p(b_1|xy_1) \, \kb{b_1} \otimes \rho_{x,y_1, b_1},
\end{equation}
where $\kb{b_1}$ is the quantum state of Roger-1, and $\rho_{x,y_1,
b_1}$ is the joint state of the rest of (unmeasured) Rogers. Note
that this map effectively measures Roger-1's part of the state,
obtains outcome $b_1$ which he stores classically, and then the rest
of Rogers are left with the state $\rho_{x,y_1, b_1}$. Now, starting
with $\rho_x$ and $\rho_{x'}$, we have that $\F(\rho_x, \rho_{x'})$
is at most $\sum_{b_1} \sqrt{p(b_1|xy_1)} \sqrt{ p(b_1|x'y_1)} \;
\F(\rho_{x, y_1, b_1}, \rho_{x', y_1, b_1})$. Note that this bound
is valid for any $y_1\in Y_1$, and thus we can take the minimum over
$y_1$, similar to the discussion after \eqref{eq:trivial}.

We can now continue this argument for each subsequent measurement
one at a time. In the second step, we consider Roger-2's local
measurement $y_2\in Y_2$ on $\rho_{x, y_1, b_1}$ and $\rho_{x',
y_1, b_1}$, which results in
\begin{widetext}
\begin{equation} \label{wide}
\F(\rho_{x, y_1, b_1}, \rho_{x', y_1, b_1}) \leq
\min_{y_2}\sum_{b_2} \sqrt{p(b_2|b_1xy_1y_2)}\sqrt{
p(b_2|b_1x'y_1y_2)} \cdot \F(\rho_{x, y_1, b_1,y_2,b_2}, \rho_{x',
y_1, b_1,y_2,b_2}),
\end{equation}
\end{widetext}
where we similarly define $\rho_{x, y_1, b_1,y_2,b_2}$ as the
quantum state of the other $k-2$ Rogers after Roger-1 and Roger-2
perform POVMs $y_1$ and $y_2$, and get outcomes $b_1$ and $b_2$
respectively. Note that in \eqref{wide} we included the minimization
over $y_2$ explicitly. Continuing further in this manner, we
eventually end up with the entire state being measured, and are left
with the relation that
\begin{widetext}
\begin{equation} \label{wide2}
\F(\rho_{x, y_1, b_1, \ldots, y_{k-1}, b_{k-1}}, \rho_{x', y_1, b_1,
\ldots, y_{k-1}, b_{k-1}}) \leq \min_{y_k}\sum_{b_k} \sqrt{ p(b_k
|b_1 \cdots b_{k-1}x y_1 \cdots y_k) } \sqrt{ p(b_k |b_1 \cdots
b_{k-1}x' y_1 \cdots y_k)}.
\end{equation}
\end{widetext}
Then by the chain rule in probability theory, we obtain the
following lemma. For simplicity, we define the vectors $\vec{b}=b_1
\cdots b_k$ and $\vec{y}=y_1 \cdots y_k$.

\smallskip
\textbf{Lemma 1.} In a prepare-and-distribute experiment generating
the correlation $p(\vec{b}|x\vec{y})$, it holds that
\begin{equation}\label{lemma}
\F(\rho_x,\rho_{x'}) \leq \displaystyle\AMS_{\vec{y}, \vec{b}} \left(
\sqrt{p(\vec{b}|x\vec{y})} \sqrt{p(\vec{b}|x'\vec{y})}
\right) ,
\end{equation}
where, for a function $f(\vec{y},\vec{b})$, we define
\begin{equation}\label{ams}
\AMS_{\vec{y}, \vec{b}} \left( f(\vec{y},\vec{b}) \right) = \min_{y_1} \sum_{b_1}
\min_{y_2} \sum_{b_2} \cdots \min_{y_k}
\sum_{b_k}f(\vec{y},\vec{b}).
\end{equation}
Here $\AMS$ is short for \emph{alternating minimization and
summation}. Note that this bound is valid for any ordering of the
Rogers, so in \eqref{ams} we also have the freedom to optimize over
such orderings.

Clearly, the bound given by the above lemma is stronger than
\eqref{eq:trivial}. Later we will see that the gap can be very
large.

\subsection{Dimension estimations in the multiparty Bell scenario}

We now turn to the main {problem} of the current paper: In a
multiparty Bell scenario, {can we test the Hilbert space dimension
of a specific party in a device-independent manner?} We designate
Alice as the party whose Hilbert space dimension we are testing and,
after fixing Alice, we may assume that the shared quantum state
$\rho$ is pure {as one of the Bobs can hold the purification of $\rho$ and
measure it trivially to obtain the same correlation data. In other words, though the result in the current subsection is proved for the case when $\rho$ is pure, it is also true for a mixed state $\rho$}.

Now let us explain the relation between multiparty Bell scenarios and
prepare-and-distribute scenarios that we mentioned earlier. Suppose Alice
measures her subsystem with any specific measurement $x$. Then
different outcomes $a$ will force the other subsystems to collapse
onto different quantum states $\rho_{x,a}$, which means she
essentially ``prepares and distributes'' $\rho_{x,a}$ on the other
subsystems with probability $p(a|x)$. Meanwhile, since the measurement on Alice's system does not affect the joint state of the other systems, for any $x$, we have that
\begin{equation}
\Tr_{\calA}(\rho) = \sum_{a} p(a|x) \rho_{x,a},
\end{equation}
where $\calA$ is Alice's Hilbert space. In this way, for any
$x,x'\in X$ we have that
\begin{equation}
\Tr(\Tr_{\calA}(\rho)^2) = \sum_{a,a'}
p(a|x) \, p(a'|x') \, \Tr(\rho_{x,a}\rho_{x',a'}).
\end{equation}
Note that
\begin{equation}
\Tr(\rho_{x,a}\rho_{x',a'})\leq\F(\rho_{x,a},\rho_{x',a'})^2.
\end{equation}
Then by Lemma 1, we have that $\Tr(\Tr_{\calA}(\rho)^2)$ is upper
bounded by
\begin{equation} \label{UB}
\sum_{a,a'} p(a|x) p(a'|x') \left( \AMS_{\vec{y}, \vec{b}}
\sqrt{p(\vec{b}|ax\vec{y})} \sqrt{p(\vec{b}|a'x'\vec{y})} \right)^2.
\end{equation}

So far, what we have done is upper bound the \emph{purity} of the
joint state of the Bobs. We now argue how this implies a dimension
bound for Alice. Since $\rho$ is pure, we have that
\begin{equation} \label{same}
\Tr( \Tr_{\calB}(\rho)^2 ) =
\Tr (\Tr_{\calA}(\rho)^2),
\end{equation}
where $\calB$ is the combined Hilbert space of all the Bobs. Since
$\Tr_{\calB}(\rho)$ is a quantum state on $\calA$, we have that
\begin{equation}\label{eq:pure}
\frac{1}{\dim(\calA)}\leq \Tr( \Tr_{\calB}(\rho)^2 ),
\end{equation}
where $\dim(\calA)$ is Alice's Hilbert space dimension. By combining
\eqref{UB}, \eqref{same}, and \eqref{eq:pure}, and using the chain
rule of probability theory
($p(a\vec{b}|x\vec{y})=p(a|x)p(\vec{b}|ax\vec{y})$), we have the
main result of this paper, below.

\smallskip
\textbf{Theorem.} In a multiparty Bell experiment generating the
correlation $p(a\vec{b}|x\vec{y})$, the Hilbert space dimension of
Alice is at least
\begin{equation} \label{thm}
\left[\min_{x,x'} \sum_{a,a'} \left( \AMS_{\vec{y}, \vec{b}} \left(
\sqrt{p(a \vec{b} | x \vec{y})} \sqrt{p(a' \vec{b} | x' \vec{y})}
\right) \right)^2\right]^{-1}.
\end{equation}
Note that the dimension of any other subsystem can be tested
similarly by defining that party to be Alice.

\begin{rem}
Note that at first glance, it seems that only one measurement of Bob is used in the bound. However, since each $y_j$ is chosen based on the measurement settings $x, x',  y_i$ and outcomes $a, a', b_i$, for $i < j$, and there is a summation over the measurement outcomes, it is likely the case that many measurement settings are used for each Bob in the computation of the bound.
If the choices of each $y_j$ were not allowed to depend on the outcomes, then one would obtain a bound much less powerful as it would not capture any of the nonlocal behaviour of the correlation.
Note that even though the measurement choices are adaptive in this regard, it does not mean we allow signalling in the experiments. This is only for the calculation of the bound (done after the experiment concludes) and does not have any physical interpretation.
\end{rem}

\subsection{Examples with tight results}

We now exhibit examples showing that the result above can perform
well. Before starting, we would like to point out that when
restricted to the bipartite case, the theorem above gives the same
result with \cite{SVV15}, which already performs very well on many
nontrivial examples of bipartite quantum correlations.

For general multipartite cases, we first show that the result
\eqref{thm} can be tight on quantum correlations with any underlying
quantum dimension. Suppose $k$ parties share a quantum state
$\sum_{i=1}^{d} \frac{1}{\sqrt{d}}\ket{i}^{\otimes k}$ and perform a
Bell experiment, where each party's measurement set includes one in
the computational basis. Suppose somehow most of the correlation
data is lost and only the part corresponding to the computational
basis measurements remain. We now use the partial data to calculate
\eqref{thm} which is  weaker than the result obtained from the full
data. However, this already proves the dimension is at least $d$,
meaning that in this case \eqref{thm} is tight, and this works for
any number of parties and any dimension $d$. Note that even though this example is rather trivial, it illustrates that our bound is not restricted in any sense to the actual minimal dimension or the number of parties involved.

Next we consider a nontrivial finite-dimensional example. The $\GHZ$
correlation is generated by the $k$-qubit quantum state
\begin{equation}
\ket{\GHZ_k} := \frac{1}{\sqrt 2} \left( \ket{0}^{\otimes k} +
\ket{1}^{\otimes k} \right),
\end{equation}
and each party has measurement settings with binary outcomes
described as the following:
\begin{eqnarray*}
\text{Pauli-}X: & & \;
\{ \ket{+}, \ket{-} \}, \\
\text{Pauli-}Y: & & \;
\{ \ket{+i}, \ket{-i} \},
\end{eqnarray*}
where
\begin{align*}
\ket{+} & = \frac{1}{\sqrt 2} (\ket{0} + \ket{1}), &
\ket{-} & = \frac{1}{\sqrt 2} (\ket{0} - \ket{1}), \\
\ket{+i} & = \frac{1}{\sqrt 2} (\ket{0} + \hat{i} \ket{1}), &
\ket{-i} & = \frac{1}{\sqrt 2} (\ket{0} - \hat{i} \ket{1}),
\end{align*}
and $\hat{i}$ is the imaginary unit. Then we have that
\begin{equation}
p(a \vec{b} | x \vec{y}) = \frac{|1+ \hat{i}^{(2 h(a \vec{b}) + h(x
\vec{y})}|^2}{2^{k+1}},
\end{equation}
where $h$ denotes the Hamming weight of a binary vector. It can be
verified that if we choose $x = 0$, $x' = 1$ and ${y_1 = \cdots =
y_{k-1} = 0}$, the lower bound for Alice's dimension is $2$ for
\emph{any} $k$, which is obviously tight. Note that in this case,
and the ones  before, the bound is exactly tight, that is, we need
not round up (noting dimension is always an integer).

The lower bound given in \eqref{thm} can also be infinite. If this
is the case, the result implies that the corresponding quantum
correlation cannot be produced by any finite-dimensional quantum
systems. For such an example, let us examine the $(k+1)$-party
PR-box \cite{PR94,BLM+05} where the correlation probability $p(a b_1
\ldots b_k | x y_1 \ldots y_k)$ can be expressed as
\begin{equation}
\frac{1}{2^{k}} \; \text{ if } \; a \oplus \bigoplus_{i=1}^k b_i =
(\Pi_{i=1}^k y_j) \cdot x, \quad 0 \text{ otherwise.}
\end{equation}
Then the bound \eqref{thm} shows that Alice's dimension must be
infinite, which can be seen as follows. We choose ${x=0}$, $x'=1$,
then when $a=a'$, let $\vec{y}$ that optimizes \eqref{thm} be
$1\cdots 1$, otherwise let it be $0\cdots 0$. This proves that this multiparty PR-Box cannot be produced by
any finite-dimensional Hilbert spaces.

\subsection{Numerical tests}
\label{SectNum}

We now assess the performance of the lower bounds given in
\eqref{thm} on tripartite quantum correlations using many examples
generated by finite-dimensional quantum systems. To produce desired
examples of quantum correlations, we fix a particular tripartite shared quantum state and generate random  measurements for Alice,
Bob, and Charlie. Specifically, {when the dimension of each local Hilbert space is $d$, each party has $d$ different measurement settings, and each of the measurements is} in the eigenbasis
of a randomly sampled symmetric matrix.
This allows us to produce many valid sets of
quantum correlation data by straightforward calculation, each of
which is generated using a finite-dimensional quantum system, where
the dimension is a tuneable parameter of our choosing. Our results
are displayed in the tables below for various choices of tripartite states.
Even though Hilbert space dimension is always an integer, we also put the exact values in the tables below.
This is done because it reveals more information about the bound, but also the exact value is relevant if the correlation is repeated many times in parallel. We see that the bound multiplies in this case, and thus the exact case is essential for this reason.

\medskip

\paragraph{\textbf{Example: High amount of entanglement}} \quad \\

The table below is for the state $\frac{1}{\sqrt{d}}\sum_{i=1}^d \ket{iii}$ on which we expect our bound to behave well due to the large amount of entanglement in the state.

\begin{table}[h]
\caption{The performance of our bound (both exact and rounded up) averaged over $100$ randomly generated tripartite quantum correlations using the state
$\frac{1}{\sqrt{d}} \sum_{i=1}^d \ket{iii}$.
Exact calculations truncated to $3$ decimal places.}
\label{Table1}
\begin{center}
\begin{tabular}{|c||c|c|c|c|}
\hline
Dimension & 2 & 3 & 4 & 5  \\ 
\hline
\hline
\; Average of (\ref{thm}) (rounded up) \; & $2.00$ & $3.00$ & $3.84$ & $4.00$ \\
\; Average of (\ref{thm}) (exact) \;          & $1.876$ & $2.538$ & $3.138$ & $3.575$ \\
\hline
\end{tabular}
\end{center}
\label{default}
\end{table}

Since each correlation is
generated using $d$-dimensional local Hilbert spaces, $d$ is a natural upper
bound on the smallest Hilbert space dimension. That being said, our lower bound performed well by
certifying this as the minimum Hilbert space dimension in most
cases. It performed near perfectly in smaller dimensions, and well in dimension $5$. Note that when testing larger
dimensional correlations, it might be the case that they are
realizable in a smaller Hilbert space dimension, thus making our
lower bound smaller in the process. On that note, it might also be
possible that our bound is performing better than we can tell,
and it is just hidden by the fact that we cannot compute the
exact minimum Hilbert space dimension. This fact is the basis of the
importance of the work in this paper.

\medskip

\paragraph{\textbf{Example: Small amount of entanglement}} \quad \\

The table below is for the state
$\sum_{i=1}^d i \cdot \ket{iii}$ (normalized) on which we expect our bound to behave less well due to the small amount of entanglement in the state.

\begin{table}[h]
\caption{The performance of our bound (both exact and rounded up) averaged over $100$ randomly generated tripartite quantum correlations using the state
$\sum_{i=1}^d i \cdot \ket{iii}$ (normalized).
Exact calculations truncated to $3$ decimal places.}
\label{Table2}
\begin{center}
\begin{tabular}{|c||c|c|c|c|}
\hline
Dimension & 2 & 3 & 4 & 5  \\ 
\hline
\hline
\; Average of (\ref{thm}) (rounded up) \; & $ 2.00$ & $ 2.00$ & $3.00$ & $3.00$ \\
\; Average of (\ref{thm}) (exact) \;          & $1.461$ & $1.865$ & $2.268$ & $2.602$ \\
\hline
\end{tabular}
\end{center}
\label{default}
\end{table}

As expected, our lower bound performs less well than the above case (which had more entanglement).
Nevertheless, it still performed decently by giving a rough estimate of $d$.
As mentioned above, it could be possible that these correlations can be generated by a quantum state of small local Hilbert space dimension.

\medskip

\paragraph{\textbf{Example: No entanglement}} \quad \\

The table below is for the mixed state
$\frac{1}{d}\sum_{i=1}^d \kb{iii}$ on which the expected success of our bound is less certain.
This is because shared randomness is not a free resource in our setting and thus even with no entanglement the bound can still be greater than $1$.

\begin{table}[h]
\caption{The performance of our bound (both exact and rounded up) averaged over $100$ randomly generated tripartite quantum correlations using the state
$\frac{1}{d} \sum_{i=1}^d \kb{iii}$.
Exact calculations truncated to $3$ decimal places.}
\label{Table3}
\begin{center}
\begin{tabular}{|c||c|c|c|c|}
\hline
Dimension & 2 & 3 & 4 & 5  \\ 
\hline
\hline
\; Average of (\ref{thm}) (rounded up) \; & $2.00$ & $2.00$ & $2.00$ & $2.00$ \\
\; Average of (\ref{thm}) (exact) \;          & $1.075$ & $1.202$ & $1.360$ & $1.451$ \\
\hline
\end{tabular}
\end{center}
\label{default}
\end{table}

\newpage

We see that our bound is rather far from $d$ for these correlations.
It is perhaps an advantage of our bound that it does not pick up Hilbert space dimension arising from shared randomness as well as it does from entanglement.
Since quantum entanglement is often viewed as
a more interesting resource than shared randomness, this advantage could be a hidden feature.

\medskip

\paragraph{\textbf{Example: Three-party Dicke state}} \quad \\

Lastly, we test our bound on the three-party Dicke state of local Hilbert space dimension $3$, as shown below:
\begin{equation}
\frac{1}{\sqrt 6} \ket{012} +
\frac{1}{\sqrt 6} \ket{021} +
\frac{1}{\sqrt 6} \ket{102} +
\frac{1}{\sqrt 6} \ket{120} +
\frac{1}{\sqrt 6} \ket{201} +
\frac{1}{\sqrt 6} \ket{210}.
\end{equation}
Below we present the numerical calculations.
\begin{table}[h]
\caption{The performance of our bound (both exact and rounded up) averaged over $100$ randomly generated tripartite quantum correlations using the state
$\frac{1}{\sqrt 6} \ket{012} +
\frac{1}{\sqrt 6} \ket{021} +
\frac{1}{\sqrt 6} \ket{102} +
\frac{1}{\sqrt 6} \ket{120} +
\frac{1}{\sqrt 6} \ket{201} +
\frac{1}{\sqrt 6} \ket{210}$.
Exact calculations truncated to $3$ decimal places.}
\label{Table4}
\begin{center}
\begin{tabular}{|c||c|}
\hline
Dimension & 3  \\
\hline
\hline
\; Average of (\ref{thm}) (rounded up) \; & $3.00$ \\
\; Average of (\ref{thm}) (exact) \; & \; $2.591$ \; \\
\hline
\end{tabular}
\end{center}
\label{default}
\end{table}

We see that this is almost the same behaviour as in  Table~\ref{Table1}, where the state tested was
\begin{equation}
\frac{1}{\sqrt 3} \ket{000} +
\frac{1}{\sqrt 3} \ket{111} +
\frac{1}{\sqrt 3} \ket{222}.
\end{equation}
The numbers suggest that, at least in the case of three parties we choose, our bound does not change greatly when the flavour of the entanglement changes in this manner.

\begin{rem}
Note that we tested hundreds of multipartite
correlations in the tables (and thousands in general) without
the need for any Bell inequalities. If we took the Bell inequality
approach, we would have to examine each correlation on its own, then
find a suitable Bell inequality that separates it from the set of
local correlations (if one even exists), then examine the extent to
which one can violate that inequality with quantum systems of
different dimensions. This is an extremely complicated and
challenging task, which we avoid entirely with our general,
easy-to-compute lower bound.
\end{rem}


\subsection{Advantage over bipartite results}

Though we are focusing on multiparty Bell scenarios in this paper,
one could in principle apply bipartite results by interpreting the
correlation as a bipartite one by combining the Bobs into a single
party. Since device-independent dimension tests already exist for
bipartite cases (for example \cite{SVV15}), this provides a simple
solution for our problem. In this situation, a natural question is
whether the new result we provide in the current paper can beat this
bipartite approach. In fact, the following example shows that
this is the case, and moreover, the advantage can be great.

Consider a three-party Bell experiment in which each party has two
binary POVMs, and the correlation is given as
\begin{equation}
\label{ex:corr} \frac{1}{4} \; \text{ if } x\cdot(y_2\oplus
b_1)=a\oplus b_1\oplus b_2, \quad 0 \text{ otherwise.}
\end{equation}
First thing we note is that this correlation is non-signalling, thus
it is conceivable that we can produce it by a quantum scheme.
Suppose this is the case, and we now focus on the dimension of
Alice's subsystem. By straightforward calculation, one can verify
that the lower bound
provided by Ref.~\cite{SVV15} is $4$, while the lower bound given by
\eqref{thm} is infinite. This means that this correlation cannot be
produced by any finite-dimensional quantum system. Clearly, this
example indicates that the result in the current paper is able to show facts that are not revealed by the bipartite results in
Ref.~\cite{SVV15}, and thus we believe is of a true multipartite
nature.

It should be pointed out that the correlation \eqref{ex:corr} is
also an example illustrating the fact that considering a different
ordering of the Bobs in our bound results in a different
performance. In fact, if we switch the roles of Bob-$1$ and Bob-$2$
the dimension bound will decay to finite.

We now perform again the numerical tests presented in Table~\ref{Table1}, but this time comparing our bound to that in Ref.~\cite{SVV15}.
See Table~\ref{Table4} below.
\begin{table}[h]
\caption{Lower bound comparison averaged over $100$ randomly generated tripartite quantum correlations using the state
$\frac{1}{\sqrt{d}} \sum_{i=1}^d \ket{iii}$ (and measurements described as in Section~\ref{SectNum}).
Exact calculations truncated to $3$ decimal places.}
\label{Table4}
\begin{center}
\begin{tabular}{|c||c|c|c|c|lc|}
\hline
Dimension & 2 & 3 & 4 & 5 \\
\hline
\hline
\; Average of (\ref{thm}) (rounded up) \; & \ $2.00$ \ & \ $3.00$ \ & \ $3.90$ \ & \ $4.00$ \ \\
\; Average of Ref.~\cite{SVV15} (rounded up) \; & $2.00$ & $3.00$ & $3.51$ & $4.00$ \\
\; Number of times (\ref{thm}) outperformed Ref.~\cite{SVV15}
\; & $0$ & $0$ & $39$ & $0$ \\
\hline
\hline
\; Average of (\ref{thm}) (exact) \; & \ $1.881$ \ & \ $2.527$ \ & \ $3.134$ \ & \ $3.590$ \ \\ 
\; Average of Ref.~\cite{SVV15} (exact) \; & $1.867$ & $2.458$ & $3.006$ & $3.435$ \\
\; Number of times (\ref{thm}) outperformed Ref.~\cite{SVV15} (by at least $0.001$)
\; & $41$ & $81$ & $92$ & $94$ \\
\hline
\end{tabular}
\end{center}
\label{default}
\end{table}

There are a few important points that the numerical results in
Table~\ref{Table4} show. Most importantly, there exist many examples
showing a finite separation between the two bounds. This illustrates
that our bound is of a true multipartite nature. These examples can
be found in dimension $4$ in the rounded case and any dimension in
the exact case. Moreover, in the exact case, we see that our bound
almost always gives a greater value. We would have liked to push
these tests further, but they get computationally expensive as the
dimension grows. On the other hand, we can already infer something
interesting even from a small gap size. As mentioned earlier, if the same correlation is
repeated many times in parallel, we see that both bounds multiply,
thus even a small gap can be amplified to arbitrarily large sizes.
Thus correlations can be constructed in this way which have
arbitrarily large finite gap.

\medskip
\subsection{Purity and entanglement test}

Going back to the proof of our theorem, we can see from
\eqref{eq:pure} that the \emph{purity} of $\Tr_{\calB}(\rho)$ is the
quantity that we actually test. Recall that the purity of a quantum
state $\sigma$ is defined as $\Tr(\sigma^2)$. It turns out that the
purity contains much more information than just a bound on the
dimension. For example, for a bipartite pure state, the purity of
reduced density matrices can be used to lower bound the amount of
entanglement. Unfortunately, in multipartite cases the situation is
much more complicated. On one hand, the concepts of entanglement
measures have not yet been fully understood for multipartite quantum
states, and on the other hand, mathematical difficulties also arise
in these cases \cite{HHHH09}. However, in our setting if somehow
more information on the structure of the shared quantum state is
already known, it is possible to draw nontrivial conclusions on
entanglement of multiparty quantum states. As an example, suppose in
addition to the correlation data, we are told that the shared
quantum state can be transferred to a state of the form
$\sum_{i=1}^ma_i\ket{i}^{\otimes n}$ by local unitary operations.
Then like in bipartite cases \cite{WS17}, we can give a nontrivial
estimation for the amount of entanglement based on only the purity
estimation of the reduced density matrices. It should be pointed out
that because of the need of extra (quantum) information, this would
no longer be fully device-independent, but still could be
interesting nonetheless, as sometimes these assumptions may be
reasonable. Rigorous device-independent techniques to test
multipartite entanglement, for example \cite{BJLP11,BSV12}, rely on
multipartite Bell inequalities that often involve complicated
geometrical characterizations of multipartite quantum correlation
sets. With these extra assumptions that we discussed, our approach
avoids such multipartite Bell inequalities which will be very
convenient for certain applications.

\section{Discussions}

In this work, we defined the prepare-and-distribute scenario, and
developed an efficient technique for estimating distances between
quantum preparations based only on measurement correlation data.
This allowed us to derive a device-independent lower bound for the
Hilbert space dimension of any given party in a multiparty Bell
scenario and gave examples showing that the result can be tight.
Furthermore, by comparing the performance of our bound with methods
based on bipartite dimension bounds, we showed that our bound is
much stronger, revealing its multipartite nature. Moreover, our
bound involves only simple functions of the correlation data, thus
being easy to calculate (all the examples in this paper can be
computed by hand), and allowing it to enjoy a robustness against
experimental uncertainty during the process of gathering the
correlation data. Considering the difficulties of generalizing dimension witnesses to multipartite cases due to the need for  multipartite Bell inequalities, we believe our approach has
great potential for future applications. In particular, like in the bipartite case (see the real-world application \cite{WPD+18}), we hope it
will prove itself useful in future quantum experiments involving
three of more parties.
\medskip
\begin{acknowledgments}
We thank Valerio Scarani and Koon Tong Goh for helpful discussions. Z.W. is supported by the
Singapore National Research Foundation under NRF RF Award
No.~NRF-NRFF2013-13, the National Key R\&D Program of China, Grant No. 2018YFA0306703, and the start-up funds of Tsinghua University, Grant No. 53330100118. Research at the Centre for Quantum Technologies
is partially funded through the Tier 3 Grant ``Random numbers from
quantum processes,'' (MOE2012-T3-1-009).
\end{acknowledgments}

\end{document}